\documentclass[aps,pre,groupedaddress,showkeys,showpacs,amsmath]{revtex4-1}
\usepackage{amssymb}
\usepackage{amsmath}
\usepackage{graphicx}
\usepackage{amsbsy}
\usepackage{bm}
\usepackage{color}
\usepackage{gensymb}
\usepackage{caption}
\usepackage{subcaption}
\usepackage[space]{grffile}
\definecolor{cream}{RGB}{222,217,201}
\usepackage{siunitx}
\DeclareSIUnit[number-unit-product = {\,}]
\cal{cal}
\DeclareSIUnit\kcal{\kilo\cal}
\DeclareSIUnit\kcal{\kilo\joule\per\mole}
\DeclareSIUnit\molar{\mole\per\cubic\deci\metre}
\DeclareSIUnit\Molar{\textsc{m}}
\usepackage{multirow}
\usepackage[version=4]{mhchem}
\usepackage{tabularx}
\newcolumntype{Y}{>{\centering\arraybackslash}X}
\usepackage{threeparttable}

\begin{document}
\title{Can Polarity-Inverted Surfactants Self-Assemble in Nonpolar Solvents?}
\author{Manuel Carrer}
\affiliation{Department of Chemistry and Hylleraas Centre for Quantum Molecular Sciences, University of Oslo, PO Box 1033 Blindern, 0315 Oslo, Norway}
\author{Tatjana \v{S}krbi\'{c}}
\affiliation{Department of Physics and Institute for Fundamental Science, University of Oregon, Eugene, OR 97403, USA}
\affiliation{Dipartimento di Scienze Molecolari e Nanosistemi, Universit\`{a} Ca' Foscari di Venezia,Campus Scientifico, Edificio Alfa, via Torino 155, 30170 Venezia Mestre, Italy}
\author{Sigbjørn Løland Bore}
\affiliation{Department of Chemistry and Hylleraas Centre for Quantum Molecular Sciences, University of Oslo, PO Box 1033 Blindern, 0315 Oslo, Norway}
\author{Giuseppe Milano}
\affiliation{Department of Organic Materials Science, Yamagata University, 4-3-16 Jonan Yonezawa, Yamagata-ken 992-8510, Japan}
\affiliation{Dipartimento di Chimica e Biologia, Universit\`a di Salerno, Via Giovanni Paolo II 132, 84084 Fisciano, Italy}
\author{Michele Cascella}
\affiliation{Department of Chemistry and Hylleraas Centre for Quantum Molecular Sciences, University of Oslo, PO Box 1033 Blindern, 0315 Oslo, Norway}
\author{Achille Giacometti}
\email{achille.giacometti@unive.it}
\affiliation{Dipartimento di Scienze Molecolari e Nanosistemi, Universit\`{a} Ca' Foscari di Venezia,Campus Scientifico, Edificio Alfa, via Torino 155, 30170 Venezia Mestre, Italy}
\affiliation{European Centre for Living Technology (ECLT) Ca' Bottacin, 3911 Dorsoduro Calle Crosera, 30123 Venice, Italy}

\date{today}

\begin{abstract}
\noindent We investigate the self-assembly process of a surfactant with inverted polarity in water and cyclohexane using both all-atom and coarse grained hybrid particle-field molecular dynamics simulations.
Unlike conventional surfactants, the molecule under study, proposed in a recent experiment, is formed by a rigid and compact hydrophobic adamantane moiety, and a long and floppy triethylene glycol tail. 
In water, we report the formation of stable inverted micelles with the adamantane heads grouping together into a hydrophobic core, and the tails forming hydrogen bonds with water.
By contrast, microsecond simulations do not provide evidence of stable micelle formation in cyclohexane.
Validating the computational results by comparison with experimental diffusion constant and small-angle X-ray scattering intensity, we show that at laboratory thermodynamic conditions the mixture resides in the supercritical region of the phase diagram, where aggregated and free surfactant states co-exist in solution.
Our simulations also provide indications about how to escape this region, to produce thermodynamically stable micellar aggregates.
\end{abstract}
\maketitle
\section{Introduction}
\label{sec:introducion}
Life as we know it could not exist without water. 
In fact, living cells survive in environments mainly constituted by water. 
Cellular shape and functionality is determined by the presence of both the plasma and the cytoplasmic membrane, which define all the necessary compartments for the organisation of the cellular matter, as well as to prevent mixing of the cell with its external environment. 
To this aim, living organisms typically exploit biological lipids, amphiphile molecules comprising a strongly polar head group and one or more long hydrocarbon tails\cite{Tanford1980}.
In aqueous solutions, these amphiphilic molecules tend to aggregate driven by 'like-to-like' interactions that are usually referred to as the hydrophobic effect\cite{Israelachvili}.
Within this general framework, water is unique because it forms hydrogen bonds with itself as well as with the polar moiety of the amphiphilic molecule.
Hydrogen bonds play a special intermediate role as they have a strength of the order of 10-40\SI{}{kJ\per\mole} (corresponding to 5-10 $k_B T/$ bond at \SI{298}{K}), much stronger than van der Waals interactions ($\approx$ \SI{1}{kJ\per\mole}),
and considerably weaker than ionic or covalent bonds ($\approx$ \SI{100}{kJ\per\mole} or more).
Also, hydrogen bonds have an intermediate orientation-dependence that is in-between the strongly directional covalent and isotropic van der Waals interactions.

While this marvelous balance is the result of millions of years of evolution, it is possible to imagine that a similar outcome could be achieved in different biological environments under different conditions, such as those present in other planets of our universe.
Although water has been detected in various thermodynamic states in our solar system, an alternative scenario suggests the possibility of using polarity-inverted membranes in non-polar solvents, such as the hydrocarbons frequently found in earth-like systems (see the recent review by Sandstr{\" o}m\cite{Sandstroem2020}).
Motivated by this idea, a number of related studies have recently been conducted.
Pace and collaborators investigated protein stability in a non-aqueous solvent such as cyclohexane (C$_6$H$_{12}$).\cite{Pace2004}
Hayashi \textit{et al.} found that while proteins have well-defined unique structures in water, this is generally not the case in other non-polar solvents.\cite{Hayashi2017, Hayashi2018}
The stability of single polar and hydrophobic amino acids in water and non-polar solvents have also been studied (unpublished results).
To close the triangle of life, the stability of B-DNA under non-aqueous conditions has also been recently assessed.\cite{Hamlin2017}

Notwithstanding the large number of studies that have been addressing the issue, the basic mechanism underlying a solvophobic effect in non-polar solvents is still far from being fully understood. 
In its simplest terms it could be stated as follows. 
If the polarity of amphiphilic molecules is inverted, so to have a hydrophobic (rather than polar) head, and a polar (rather than hydrophobic) tail, would they self-assemble in non-polar solvents such as, for example, C$_6$H$_{12}$?
And if so, what would the driving force be?

\begin{figure}[!htb]
    \centering
    \includegraphics{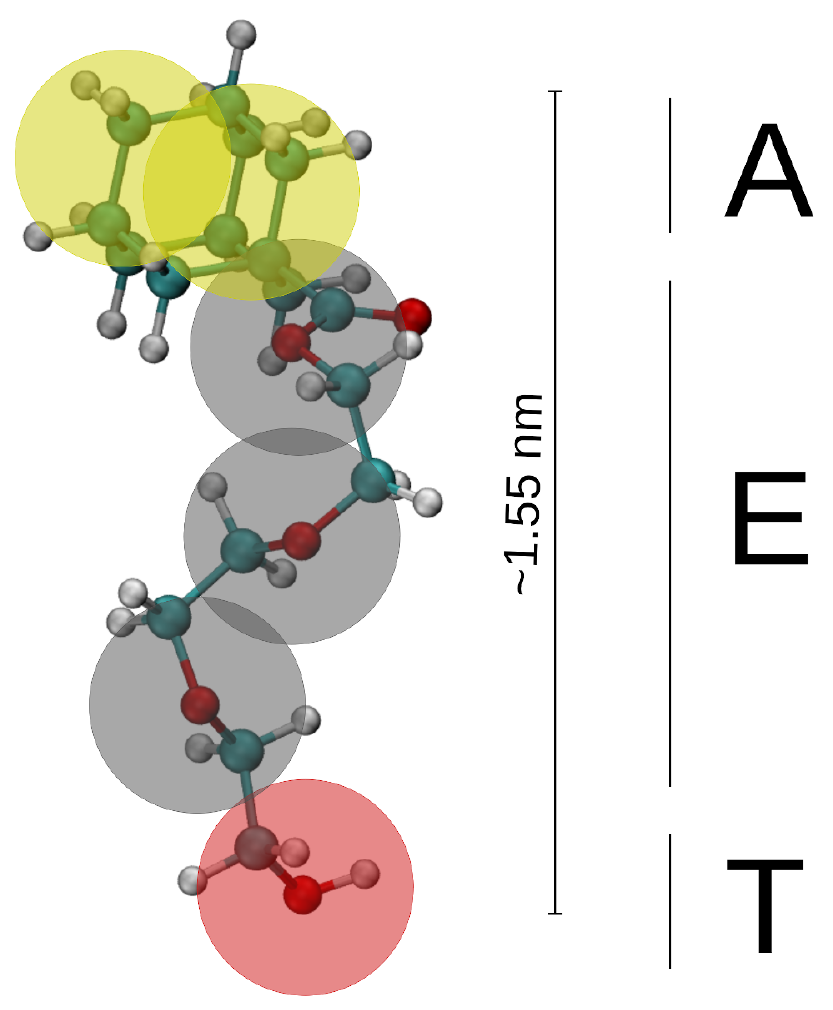}
    \caption{Molecular structure of ADOH. The transparent beads represent the coarse grained mapping used for the hPF-MD simulations. The labels identify the three different functional groups: A = adamantane, E = TEG, T = OH.}
    \label{fig:mapping}
\end{figure}

Recently such an experiment has been performed on a newly synthesized molecule having exactly these features\cite{Facchin2017}. 
This molecule, henceforth referred to as ADOH, is formed by a rigid and compact adamantane moiety \ce{AD} and a long and floppy tail that consists of a triethylene glycol (TEG) with a characteristic group \ce{O-CH2-CH2} capped at the end by a hydroxyl group that is able to form hydrogen bonds (Figure~\ref{fig:mapping}). 
The self-assembly properties of ADOH were studied by measuring its diffusion coefficient in C$_6$H$_{12}$, using nuclear magnetic resonance (NMR) spectroscopy at different concentrations.
\textcolor{black}{A monotonic decrease of the diffusion coefficient, which is a possible signal of the micellization process, was observed in the concentration range from 5 to 250 \si{mM}} and this hypothesis was further supported by small angle X-ray scattering (SAXS) measurements that appeared to indicate a critical micelle concentration (CMC) around \SI{100}{mM}.
While through the Stokes-Einstein equation this decrease in the diffusion coefficient can be ascribed to the appearance of large aggregates, it alone does not constitute proof of the existence of aggregates with well-defined micellar shape, especially considering the fact that directional-dependent polar interactions, such as hydrogen bonds, are significantly stronger compared with non-directional van der Waals interactions.
Moreover, while SAXS experiments provide essential system information on properties such as micelle sizes, the interpretation of aggregates in terms of shapes requires post-processing by modeling, which is prone to errors. 
Thus a molecular picture of aggregation is often hard to obtain.  

Molecular modelling can complement experiments by providing molecular resolution predictions on the spatial organization of the molecules. 
\textcolor{black}{}In general, surfactant aggregation is a challenging process to simulate as it is facilitated by slow diffusing molecules and typically occurs at very low concentrations requiring both very large system sizes and long simulation times.
For this specific ADOH surfactant system, the expected drop in diffusion coefficient occurs at very high concentrations, making it within reach of standard all-atom simulations. 
Nevertheless, all-atom simulations are computationally expensive and, even for at this high concentration regime, they may be affected by significant finite 
size effects and may be limited in the range of accessible time scales.

Both of these drawbacks can be effectively addressed using coarse-grained modeling\cite{voth2008coarse,Marrink2013}.
In coarse-grained models, a lower resolution representation of the molecular structure with effective potentials is used to lower the overall computational cost, thereby allowing for the study of larger and longer simulations. 
For soft matter systems in particular, such as ADOH, the coarse-grained methodology of hybrid particle-field molecular dynamics (hPF-MD) has already been proved to be particularly effective.
In hPF-MD the coarse-grained molecular resolution description is combined with density-field modeling of intermolecular interactions to yield a computationally efficient modeling of very large systems.
Applications of hPF-MD have started from more conventional soft polymer mixtures and then moved also to biological systems~\cite{Milano2013PHYSBIO,Soares2017JPCL,Cascella2015CHEMMOD,Marrink2019CHEMREV}.
Examples from the literature include nanocomposites, nanoparticles, percolation phenomena in carbon nanotubes~\cite{Nicola2016EPJ,Zhao2016NANOSCALE,Munao2018NANOSCALE,Munao2019MACRO}, lamellar and nonlamellar phases of phospholipids~\cite{Nicolia2012TCA,ledum2020automated,Nicolia2011JCTC,Bore2020JCP}, and more recently polypeptides~\cite{bore2018hybrid}, and  polyelectrolytes~\cite{Zhu2016PCCP,kolli2018JCTC,Bore2019JCTC,Denicola2020BBA,Schaefer2020}. 
These applications, in particular polyelectrolytic molecules and surfactants (Triton X-100)\cite{Pizzorusso2017}, give us sufficient ingredients for building a hPF-MD model for ADOH. 

Thus, the aim of the present work is to provide a molecular understanding on the nature of the putative ADOH aggregates reported in Ref.~\cite{Facchin2017}, as well as the underlying physical driving forces.
Using both all-atom and hPF-MD numerical simulations, the self-assembly properties of ADOH molecules will be studied both in C$_6$H$_{12}$ and in water, under the same conditions reported in the experiment.
While for the former case this will provide a complementary description with respect to the experiment, the latter case in water represents a new prediction that could eventually lead to further experimental testing.

The rest of the paper is organized as follows.
Section \ref{sec:methods} outlines the all-atoms and hPF-MD methods used in the present paper.
Section \ref{sec:results} reports the results while Section \ref{subsec:comparison} provides a connection to the experimental findings.
Finally, Section \ref{sec:conclusions} includes the key messages of the present study as well as some perspectives for future work.
\section{Methods}
\label{sec:methods}
\subsection{All-atom simulations}
\label{subsec:atomistic}
$NPT$ all-atom simulations were run using the OPLS-AA force field\cite{oplsaa} with a time step of \SI{2}{fs}.
The temperature was set to \SI{300}{K} and the pressure to \SI{1}{atm}. 
The coupling was ensured by applying the v-rescale thermostat\cite{v_rescale}, with a relaxation time of \SI{0.1}{ps}, and the Parrinello-Rahman barostat\cite{barostat}, with coupling constant set to \SI{3}{ps} and isothermal compressibility equal to \SI{4.5e-5}{\per\bar}.
Long-range electrostatics was calculated with the PME method, using a fourth order interpolation, a \SI{0.16}{nm} Fourier spacing and a \SI{1.2}{nm} cutoff, which was the same also for the calculation of short-range van der Waals interactions. 
Bond lengths were constrained using the LINCS algorithm\cite{lincs}.
The water model used was the TIP4P\cite{Jorgensen_tip4p}, while for C$_6$H$_{12}$ we used the parameterization implemented in the OPLS-AA force field\cite{Jorgensen1996_chex}.
The all-atom simulations were run with the GROMACS 2018.4 software\cite{gromacs}.

\subsection{The hPF-MD approach}
\label{subsec:hybrid}
In hPF-MD, molecular dynamics is used to sample the phase space of \textcolor{black}{a fully-resolved molecular system composed by $N_{mol}$ molecules} with Hamiltonian
\begin{equation}
    \mathcal{H} = \sum_{n=1}^{N_{mol}} \mathcal{H}_0 (\mathbf{r}_n) + \mathcal{W} [\phi(\mathbf{r})].
\end{equation}
Here $\mathcal{H}_0 (\mathbf{r}_n)$ is the Hamiltonian of a single non-interacting molecule and $\mathcal{W} [\phi (\mathbf{r})]$ is the interaction energy that depends on the particle density $\phi$. 

To model non-bonded attraction and repulsion between particles, we employ the following interaction energy\cite{Milano2009}
\begin{equation}
    \mathcal{W} [\phi (\mathbf{r})] = \frac{1}{2\phi_0} \int \text d\mathbf{r}~\left[ \sum_{ij} \tilde \chi_{ij} \phi_i(\mathbf{r}) \phi_j (\mathbf{r}) + \frac{1}{\kappa} \left(\sum_{i} \phi_i (\mathbf{r}) - \phi_0 \right)^2\right],
\end{equation}
where $\phi_0$ is the total number density, $\tilde \chi_{ij}$ is the interaction term between species $i$ and $j$, $\phi_i(\mathbf{r})$ and $\phi_j(\mathbf{r})$ are the number densities of the $i$th and $j$th species calculated at positions $\mathbf{r}$, and $\kappa$ is a compressibility term.
The net effect of $\mathcal W$ is an external potential $V_i$ \textcolor{black}{acting on all particles of type $i$}, which is obtained by
\begin{equation}
    V_i (\mathbf{r}) = \frac{\delta\mathcal{W} [\phi(\mathbf{r})]}{\delta \phi_i (\mathbf{r})}
    = \frac{1}{\phi_0} \left[ \sum_{j} \tilde \chi_{ij} \phi_j (\mathbf{r}) + \frac{1}{\kappa} \sum_{j} \left(\phi_j (\mathbf{r}) - \phi_0\right) \right].
\end{equation}
\textcolor{black}{The force acting on a particle of type $i$ is obtained by gradient operation on V: 
\begin{equation}
    \mathbf{F}_i (\mathbf{r}) = - \bm{\nabla}V_i = -\frac{1}{\phi_0}  \sum_{j}\left( \tilde \chi_{ij} + \frac{1}{\kappa} \right) \bm{\nabla}\phi_j (\mathbf{r}).
\end{equation} 
}
Calculation of the potentials and of the forces acting on the particles, and used to integrate the equations of motion, are computed with a particle mesh approach. For more details we refer to Ref\cite{Zhao2012JCP}.

\subsection{hPF-MD simulations}
Figure~\ref{fig:mapping} shows the CG mapping chosen for ADOH, while in Table~\ref{tab:ADOH_chi} we report the bead interaction matrix $\tilde \chi_{ij}$ used in the present work.
The $\tilde \chi$ values for C$_6$H$_{12}$ and ADOH were \textcolor{black}{selected} from chemically similar moieties of Triton X-100 from Ref\cite{Pizzorusso2017}.

\begin{table}[!htb] 
	\centering
	\begin{tabularx}{\textwidth}{c|Y Y Y Y Y }
		\hline\hline
					    & A 	& E 	& T 	& C (\ce{C6H12})	& W (\ce{H2O})\\\hline
		A				& 0 	& 7.8 	& 13.25 & 0 				& 33.75		\\
		E				& 7.8	& 0		& 4.5 	& 7.8 				& 1.5		\\
		T				& 13.25 & 4.5	& 0		& 13.25				& 0			\\
	    C (\ce{C6H12})& 0	& 7.8	& 13.25 & 0				    & ---	\\
		W (\ce{H2O})		& 33.75	& 1.5	& 0		& ---				& 0			\\
		\hline\hline
	\end{tabularx}
	\caption{Interaction matrix $\tilde \chi_{ij}/\si{kJ.mol^{-1}}$ used in the hPF-MD simulations.}
	\label{tab:ADOH_chi}
\end{table}
CG simulations were run in the $NVT$ ensemble using the OCCAM software\cite{Zhao2012JCP}.
The time step was set to \SI{30}{fs} and the temperature was kept constant at \SI{300}{K} by applying the Anderson thermostat\cite{Andersen1980} with collision frequency of \SI{7}{\per\pico\s}.
The density field was updated every 20 time steps.
In this model each C$_6$H$_{12}$ molecule is represented by a single bead, while a water bead comprises four molecules.
\textcolor{black}{For the simulations in \ce{C6H12} we added small alternating partial charges (+0.4 and -0.4, chosen equal as the all-atom charges assigned to ether oxygens) on ADOH tail beads.
This was done in order to qualitatively mimic the weak electrostatic interactions that, from atomistic simulations of the system with and without partial charges on the TEG segment, were found crucial in describing the clustering of the surfactant in the apolar solvent.}
The electrostatic potential in this case follows the Poisson equation with a relative dielectric constant $\epsilon_r=$ 5 and its computation is performed using the recently developed adapted PME-approach as in references\cite{Zhu2016PCCP,kolli2018JCTC}.
The composition of all simulated systems is reported in Table~\ref{tab:simulation_details}.

\begin{table} 
	\centering
	\begin{threeparttable}
	\begin{tabularx}{\textwidth}{c|YYYYYY}
		\hline\hline
		Concentration	& Method		& Solvent  	& ADOH molecules	& Solvent molecules	& Box size (\si{nm})	& Time (\si{ns})	\\\hline
		1 mM    & hPF-MD		& \ce{H2O}		& 13  		& 182801        	& 28			& 100		\\
		2 mM    & hPF-MD		& \ce{H2O}		& 26  		& 182704			& 28			& 100		\\
		5 mM    & hPF-MD		& \ce{H2O}		& 66  		& 182412	    	& 28			& 100		\\
		10 mM 	& hPF-MD		& \ce{H2O}		& 132  		& 181927			& 28			& 100		\\
		15 mM   & hPF-MD		&\ce{H2O} 		& 198  		& 181441        	& 28			& 100		\\
		50 mM 	& hPF-MD		&\ce{H2O} 		& 661  		& 178041			& 28			& 100		\\
		50 mM 	& All-atom*	    &\ce{H2O}  		& 661  		& 712490 			& 28			& 50		\\
		100 mM 	& hPF-MD		&\ce{H2O}  		& 1322 		& 173184			& 28			& 100		\\
		200 mM 	& hPF-MD		&\ce{H2O}  		& 2644 		& 163469			& 28			& 100		\\
		200 mM 	& hPF-MD		& \ce{C6H12}    & 330  		& 13670				& 14			& 1800		\\
        200 mM	& All-atom*	    & \ce{C6H12}    & 330  		& 11671				& 14			& 150		\\
		200 mM	& All-atom	    & \ce{C6H12}   & 330  		& 11079				& 14			& 450		\\\hline\hline
	\end{tabularx}
	\begin{tablenotes}
	    \small
	    \item * Model with only an effective charge in the terminal hydroxyl group of the surfactant tail.
	\end{tablenotes}
	\end{threeparttable}
	\caption{Simulation setup.}
    \label{tab:simulation_details}
\end{table}

\section{Results and Discussion}
\label{sec:results}
\subsection{ADOH in cyclohexane}
\label{subsec:cyclohexane}
We start our analysis by presenting simulation results for ADOH in C$_6$H$_{12}$, i.e. the same system studied experimentally in Ref.\cite{Facchin2017}.
In this experiment, results from SAXS and NMR suggested the onset of micelle formation, with a CMC of approximately \SI{100}{mM}, and an average radius estimated between \SI{1.7}{nm} and \SI{2.5}{nm}.
As the contour length of ADOH is $\approx$ \SI{1.5}{nm}, the emerging scenario was that of nearly spherical micelles with all of the TEG tails buried deeply inside each micelle and the adamantane hydrocarbon heads in contact with the solvent.
Hence, the analogue of a conventional spherical micelle with reversed polarity of \textit{both} the surfactant \textit{and} the solvent.

Spherical micelles can indeed be expected on the basis of the surfactant packing parameter \cite{Israelachvili1976}
\begin{equation}
    N_s = \frac{V_{t}}{A_c \ell_t},
\end{equation}{}
where $V_t$ and $\ell_t$ are the tail volume and length respectively, and $A_c$ is the contact area between the head group and the tail.
In the case of ADOH $\ell_t =0.93-1.25$ \si{nm}, $V_t \approx$  \SI{0.603}{nm\cubed}, and $A_c \approx$  \SI{1.55}{nm\squared}, as obtained by approximating the head group to a sphere and assuming the contact area as half of the sphere surface. 
This leads to $N_s=0.312-0.421$, consistent with a surfactant forming spherical or cylindrical aggregates.
We remark, however, that this argument assumes a tight packing in a straight conformation of the TEG tails into the core of the micelle, which contradicts the characteristic high flexibility associated to the chemical structure of the tail.

We first consider all-atom simulations in order to have full control over the driving forces at the microscopic level.
\textcolor{black}{The simulations were run at \SI{200}{mM}, a concentration significantly higher than the putative experimental CMC at which an increase of the SAXS intensity was observed ($\approx$ \SI{100}{mM}).}

\begin{figure} 
    \centering
    \includegraphics{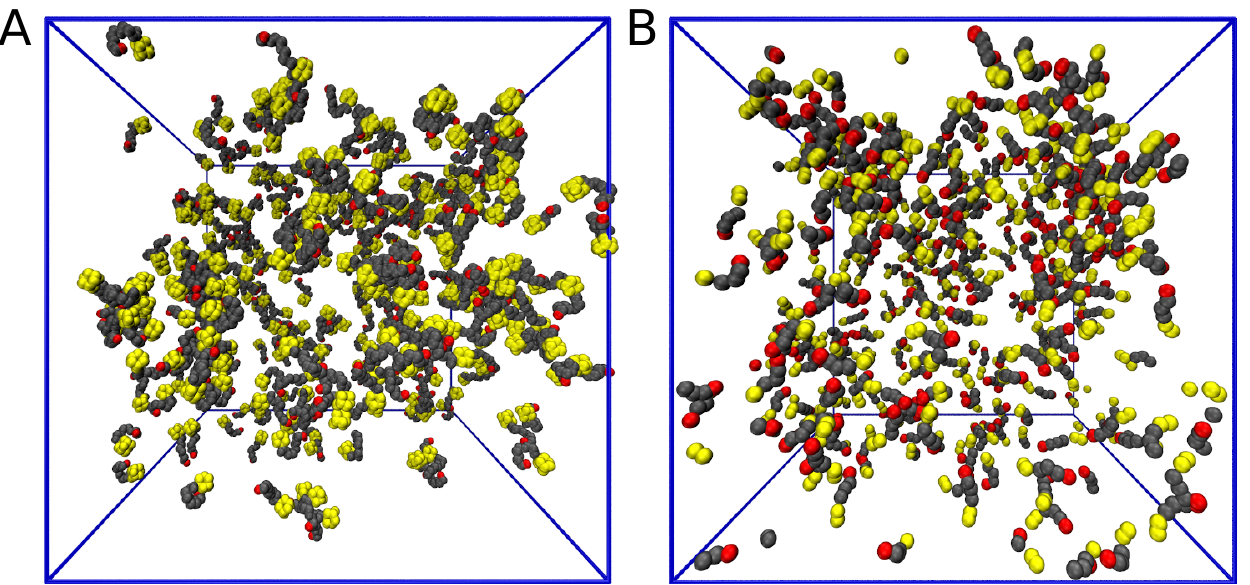}
    \caption{Final snapshots from (A) all-atom after \SI{450}{ns} and (B) hPF-MD after \SI{1800}{ns} simulations of 200 mM ADOH in C$_6$H$_{12}$. In both cases, the box size is \SI{14}{nm}. Solvent molecules have been removed for clarity.}
    \label{fig:chex_final}
\end{figure}

\begin{figure}[h]
    \centering
    \includegraphics{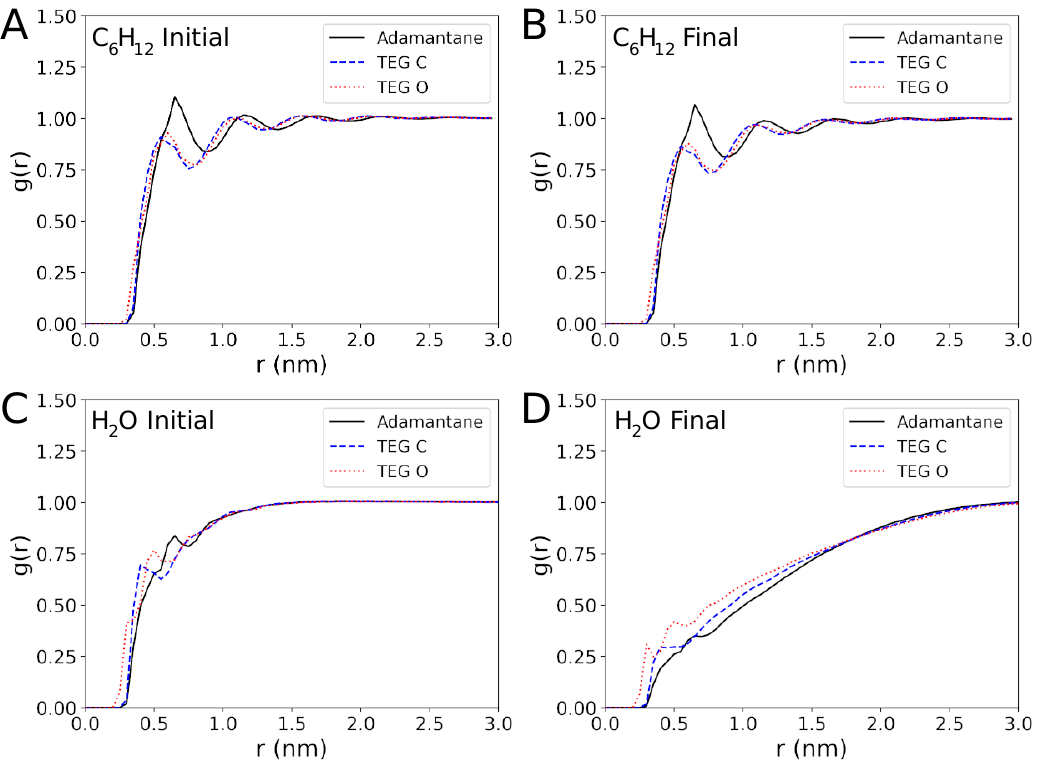}
    \caption{RDF between the solvent and the adamantane head (solid black line), TEG carbons (dashed blue line) and TEG oxygens (dotted red line) at the start and end of the simulation for both \ce{C6H12} (A, B) and water (C, D).}
    \label{fig:solvent_rdf}
\end{figure}

Figure \ref{fig:chex_final}A shows a snapshot of the final configuration after \SI{450}{ns}, and no clear sign of any sort of aggregation is visible.
Clearly, this may be an issue of the atomistic simulation timescale. Indeed, even  in conventional surfactants although early aggregation already occurs in the first nanoseconds of simulation, the timescales for the stabilization of micelles usually extends to the microseconds regime, and can be most suitably probed by coarse-grained models, such as hPF-MD.
The absence of stable aggregates is however confirmed by hPF-MD, as depicted by the corresponding snapshot of Figure \ref{fig:chex_final}B obtained after \SI{1800}{ns}.
{}
\begin{figure}
	\centering
    \includegraphics{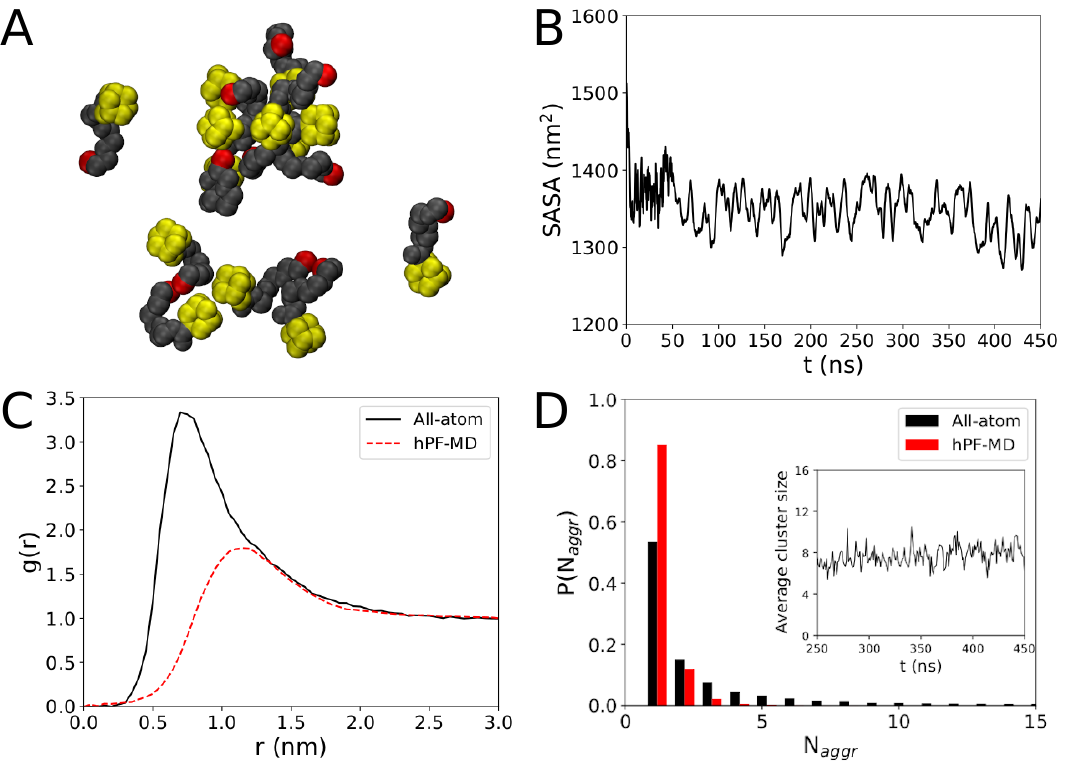}
	\caption{(A) A zoomed-in picture from the last frame of the all-atom simulation. (B) All-atom solvent accessible surface area. (C) ADOH-ADOH radial distribution function $g(r)$ from both all-atom and hPF-MD simulations. (D) Cluster size distribution of both all-atoms and hPF-MD simulations. The inset depicts the average cluster size as a function of time.}
	\label{fig:gromacs_chex_struct}
\end{figure}
Zooming in on the all-atom snapshot of Figure \ref{fig:chex_final}A, it is possible to see a seven molecule aggregate, two dimers kept together by tail-tail interactions of the hydroxyl groups, and two free monomers (Figure \ref{fig:gromacs_chex_struct}A). 
This represents a typical transient molecular cluster that is frequently observed during the simulation, characterised by irregular shape and very short lifetime, in the $10^1-10^2$~ps range.
The absence of stable micelles at this concentration is confirmed by the practically flat time profile of the solvent accessible surface area (SASA) obtained from all-atom simulations (Figure~\ref{fig:gromacs_chex_struct}B) indicating the absence of a core collapse.
The dynamical behavior of the system (see movie provided in SI) clearly indicates the presence of fast forming/disrupting dimers, trimers and higher order multimers.
Nonetheless, aggregation does not seem to result in the formation of a typical core, ADOH diffusion appears dominated by the monomeric phase, while the nonpolar solvent does not show any preferential affinity for either the head or the tail group. 
In particular, the first peak of the radial distribution function (RDF) of \ce{C6H12} against the adamantane head or the TEG tail atoms is found for both cases at $\sim$\SI{0.6}{nm}, while the profile shows the characteristic modulations typical of a good solvent (Figure~\ref{fig:solvent_rdf}A,~\ref{fig:solvent_rdf}B).
Moreover, the RDF plot does not change over time, indicating the absence of progressive aggregation during the simulation.
The all-atom ADOH-ADOH center of mass RDF reported in Figure~\ref{fig:gromacs_chex_struct}C shows a broad peak at $\approx$\SI{0.75}{nm} and tends to reach unity at a distance of $r \approx $ \SI{2}{nm}, which is an indication of the presence of uncorrelated disorder beyond this range.
Also shown in Figure~\ref{fig:gromacs_chex_struct}C the RDF from hPF-MD calculation that provides a consistent picture although with a peak at a slightly larger value indicating the tendency for the ADOH molecules to settle at this distance on a longer time scale.  

The absence of well-defined aggregates for ADOH in C$_6$H$_{12}$ is finally confirmed by the cluster size distribution (Figure~\ref{fig:gromacs_chex_struct}D), which exhibits an exponential decay with cluster size $N_{\text{aggr}}$.
The average cluster size remains constant during the simulation around a value of $\approx$\SI{7}{}, indicating that a steady state has been reached.
This is further corroborated by hPF-MD simulations that do not show any formation of stable micelles even after \SI{1800}{ns} (see Figure~\ref{fig:chex_final}B) and that show the same exponential decay found for the all-atom case in the aggregation number analysis.

\subsection{ADOH in Water}
\label{subsec:water}
\begin{figure} 
    \centering
    \includegraphics{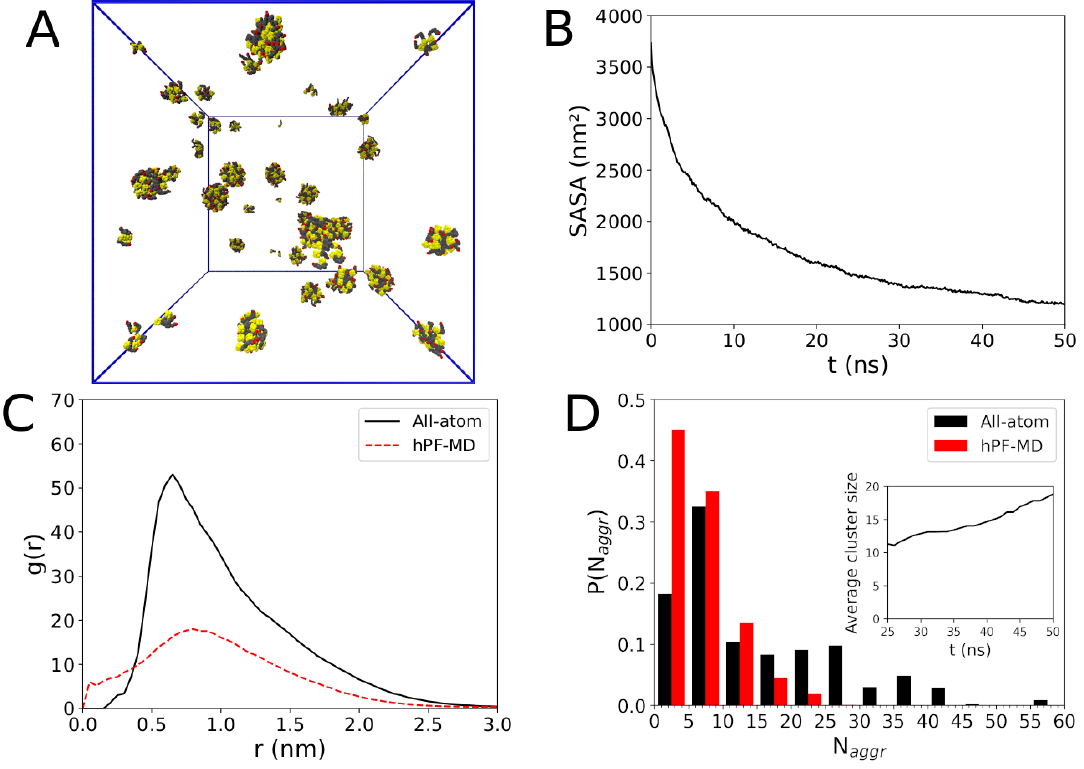}
    \caption{(A) Snapshot of a 50 mM ADOH solution in water after \SI{50}{ns} of all-atom MD simulations. Water molecules have been removed for clarity; (B) All-atoms solvent accessible surface area as a function of time; (C) ADOH-ADOH radial distribution function; (D) cluster size distribution. The inset shows the average cluster size as a function of the simulation time.}
    \label{fig:aa_water}
\end{figure}

To better understand the physics of ADOH in solution, it proves interesting to investigate the self-assembly of ADOH in water, a case that unfortunately was not covered in the experiment of Ref.\cite{Facchin2017}.  
Figure \ref{fig:aa_water}A depicts the aggregation state for a \SI{50}{mM} concentration of ADOH in water after only \SI{50}{ns} of all-atom MD simulations.
{}
Even a very short simulation time window is sufficient to report evidence that ADOH self-assembles in regular spherical micelles, in agreement with the predictions given by its packing parameter of 0.312 and its simple molecular structure (Figure \ref{fig:mapping}).
The bulky hydrophobic adamantane heads of the amphiphilic molecule promote fast aggregation to minimize the contact with water, grouping together into a well recognizable hydrophobic core, while the hydroxyl groups of the TEG tails stick out into the solvent. 
This is also well depicted by the time evolution of the RDF profile between the ADOH moieties and the solvent (Figure~\ref{fig:solvent_rdf}C,~\ref{fig:solvent_rdf}D).
In particular, at the end of the simulation, RDF profiles reach bulk values at much longer distances than in \ce{C6H12}, a consequence of the micellar collapse that screens ADOH moieites from the solvent. 
There is also a clear differentiation between the RDF profiles of TEG oxygens and adamantane, indicating the expected clear preference of the solvent for the former.
The opposite behaviour is instead not occurring in \ce{C6H12}.

The formation of micelles at this concentration is confirmed by the analysis of the SASA that, unlike in \ce{C6H12}, steadily decreases and levels off toward stable value already after \SI{50}{ns} of MD simulations (Figure~\ref{fig:aa_water}B).
This is further supported by the ADOH-ADOH center of mass RDF reported in Figure~\ref{fig:aa_water}C.
A very broad peak at $\approx$\SI{0.75}{nm} indicates the smallest ADOH-ADOH distance in a dimer (see Figure~\ref{fig:aa_water}A), while a slow decaying signal that asymptotically reaches unity at \SI{3}{nm}, is indicative of the several additional characteristic ADOH-ADOH distances present in the micelle, all consistent signs of the presence of stable aggregates.
\textcolor{black}{The cluster size distribution analysis for the micelles (Figure \ref{fig:aa_water}D, here the bin size is five units) corroborates this picture with the visible presence of a multivalued distribution and a relatively large polydispersity in micelle size, which is mainly due to the relatively short simulation times.}
In fact, the plot of the average cluster size as a function of time is characterised by a gradual increase in the value, indicating that the final equilibrium state is not reached.

In any case, the all-atom results are sufficient to indicate that the simulated system is well above the critical micelle concentration (CMC).
To obtain a rough estimate of the CMC of ADOH in water, we repeated simulations of ADOH at progressively lower concentrations using a hPF-MD approach.
Thanks to the cheaper potentials and the intrinsically accelerated dynamics, hPF-MD yields a better convergence of the aggregation state of ADOH even at much lower concentrations than 50~mM.

\begin{figure} 
    \centering
    \includegraphics{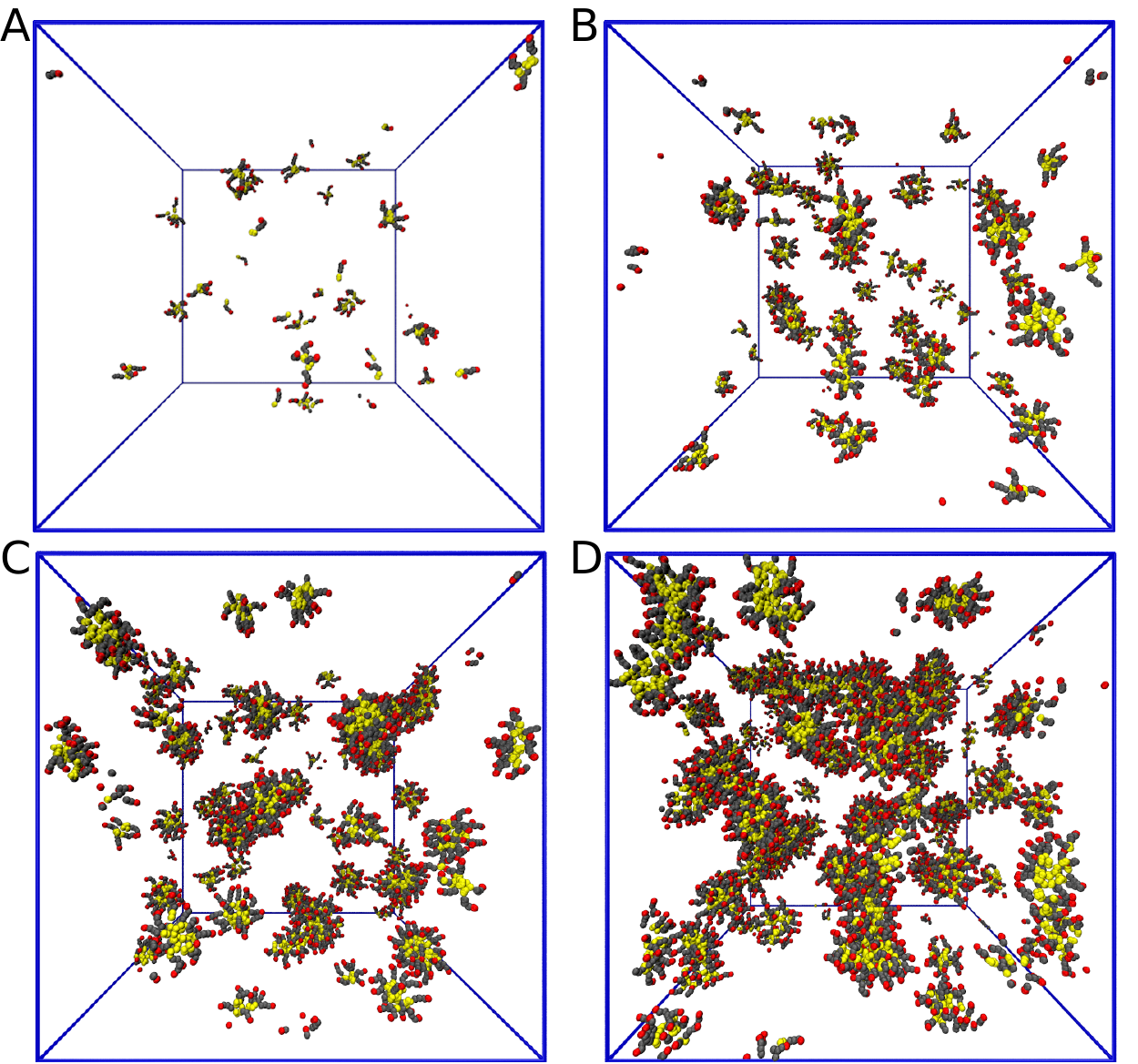}
    \caption{Hybrid particle field snapshots of micelles formed after \SI{100}{ns} in a \SI{28}{nm} cubic box at different concentrations: (A) \SI{10}{mM}, (B) \SI{50}{mM}, (C) \SI{100}{mM} and (D) \SI{200}{mM}.}
    \label{fig:hpf_water}
\end{figure}

The hPF-MD model was first validated against reference all-atom data, by running simulations at 50~mM.
Figure \ref{fig:hpf_water} presents snapshots of the final configurations obtained after \SI{100}{ns} of hPF-MD simulations at \SI{10}{mM}, \SI{50}{mM}, \SI{100}{mM}, and \SI{200}{mM} concentrations. 
These snapshots provide clear indications of aggregation, also supported by the corresponding ADOH-ADOH RDF (not shown), all exhibiting a peak localised at 0.75 nm and a similar long-range decay (see Figure \ref{fig:aa_water}C).
Note that the discrepancy on the height of peak and higher distribution values at short range are a typical feature of hPF models, and must be attributed to the soft nature of the hPF potential.

{}
From the morphological viewpoint, the size of the aggregates increases with solute concentration and at \SI{200}{mM} micelles start fusing together to form tubular structures.
This is consistent with the concentration-dependent smooth transition from spherical to tubular micelles that is commonly observed in more conventional amphiphilic surfactants, and can be ascribed to the reaching of a critical packing of the hydrophobic moieties above which they can no longer be accommodated into a compact spherical volume.\cite{Velinova2011,DeNicola2015,Pizzorusso2017}

\begin{figure}
    \centering
    \includegraphics{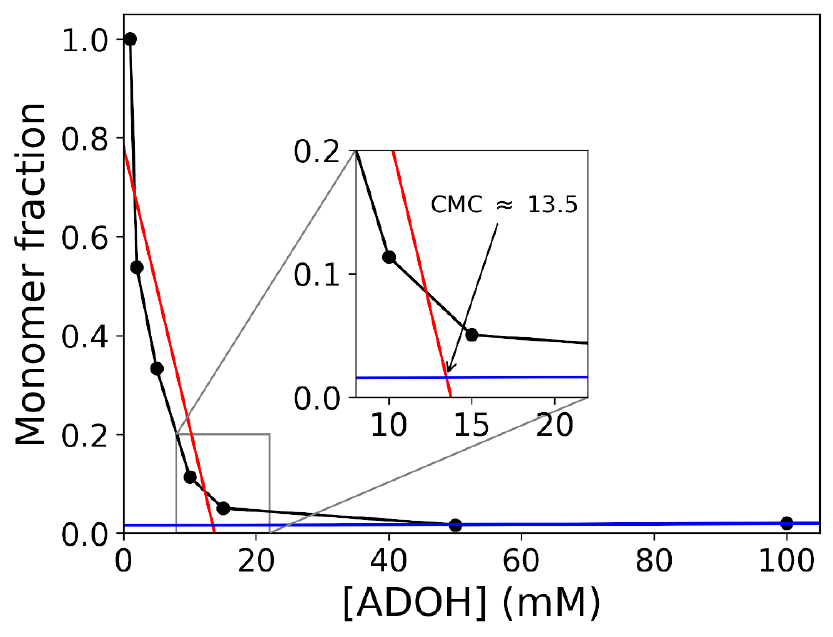}
    \caption{CMC determination by linear fitting of the two distinct regions present in the monomer fraction vs surfactant concentration plot.}
    \label{fig:cmc}
\end{figure}

The absence of significant aggregated units at \SI{10}{mM} suggests that, in water, the CMC for ADOH is between \SI{10}{mM} and \SI{50}{mM}.
We repeated additional hPF-MD simulations at progressively higher ADOH dilutions (\SIlist{1;2;5;15}{mM}).
By performing a linear fit on the concentration dependence of the free monomer fraction on the surfactant concentration at smaller and higher values than the CMC, and determining their intersection point, we produce a best estimate for the CMC $\approx$ \SI{13.5}{mM} (Figure~\ref{fig:cmc}).
\textcolor{black}{Here we notice that the number of surfactant molecules present in the lowest concentration simulation is quite small so the value of the first point in Figure~\ref{fig:cmc} could be lower if we accounted for possible finite size errors.
However even changing the monomer fraction in the range 0.9-0.7, the error on the CMC is approximately only of the $\sim$ 1-4\%.}

{}
\subsection{Reconciling with the experiment}
\label{subsec:comparison}
The interpretation of the experimental results of ADOH in C$_6$H$_{12}$ in terms of formation of well defined micelles above a CMC $\approx$\SI{100}{mM}\cite{Facchin2017} and the numerical simulations that found no evidence of a stable micellization process, are in apparent contradiction.

In fact, formation of aggregates was originally suggested on the basis of two experimental evidences: {\it (i)} a significantly lower diffusion constant for ADOH compared to that of the pure solvent, and {\it (ii)} a marked change in the SAXS intensity signal for concentrations above 100~mM.

\paragraph{Diffusion coefficient} Diffusion coefficients for C$_6$H$_{12}$, water and ADOH in the two solvents were estimated from the slope of the mean square deviation (MSD), $\sigma^2(t)$, calculated for all-atom simulations according to the Einstein diffusion equation $\sigma^2(t) = 6Dt + A$, and compared with experimental results obtained from $^1$H 2D-DOSY NMR measurements.
Table~\ref{tab:diffusion} reports the comparison between experimental diffusion coefficients (first column) and the corresponding all-atoms estimates (third column), and provides evidence of a reasonably good agreement. 
The results of the second column will be discussed further below.
On this basis, we now argue that the drop observed in the experiment might be due to other reasons and not related to the micellization process.
\begin{table}[] 
	\centering
		\begin{tabular}{c|ccc}
			\hline\hline
			& Experimental & All-atom w/o partial charges & All-atom w/ partial charges \\\hline
			\ce{C6H12} & 1.48 $^{\text{Ref.~\cite{Facchin2017}}}$ & 1.50 $\pm$ 0.02 & 1.38 $\pm$ 0.02 \\
			ADOH & 0.47 $^{\text{Ref.~\cite{Facchin2017}}}$ & 0.63 $\pm$ 0.03 & 0.48 $\pm$ 0.01 \\\hline
			\ce{H2O} & 2.30 $^{\text{Ref.~\cite{Holz2000}}}$ & 3.90 $\pm$ 0.01 & --- \\
			ADOH   & --- & 0.35 $\pm$ 0.02 & --- \\\hline\hline
		\end{tabular}{}
	\caption{\textcolor{black}{Diffusion coefficients (\SI{e-5}{cm\squared\per\second}) calculated from mean square displacement analysis of the all-atom simulations compared with experimental results. First column, experimental results (250 mM); Second column all-atoms simulations without partial charges; Third column all-atoms simulations with partial charges. The concentrations are 200 mM for simulations in \ce{C6H12} and 50 mM for \ce{H2O}.}}
	\label{tab:diffusion}
\end{table}

\begin{figure}
    \centering
    \includegraphics{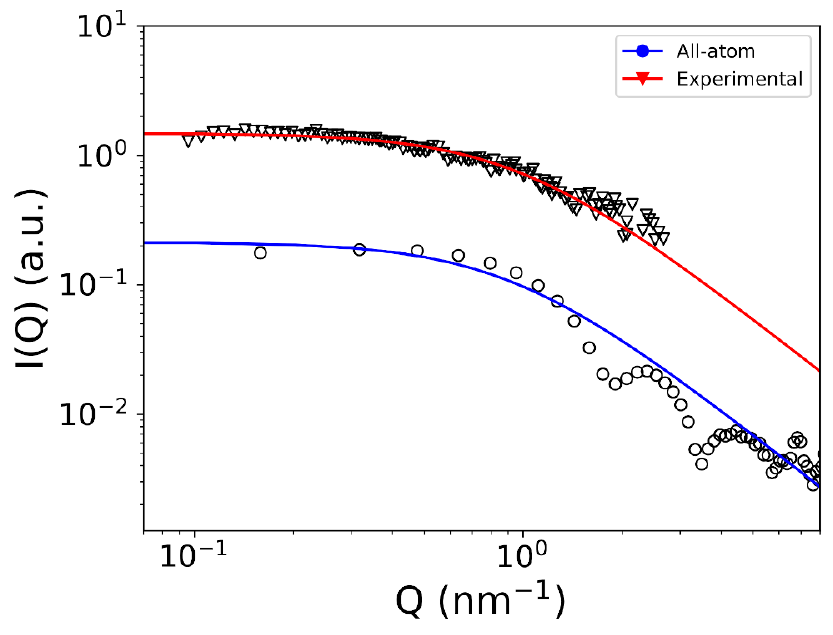}
    \caption{SAXS data extracted from Ref.~\cite{Facchin2017} (triangles) compared to the one calculated by a Fourier transform of the RDF from our all-atom simulation at 200 mM ADOH in \ce{C6H12} (circles). The red and blue lines are the corresponding fits to Equation \ref{eq:intensity_saxs}.}
    \label{fig:saxs_fit}
\end{figure}

\paragraph{SAXS Spectrum}
Figure~\ref{fig:saxs_fit} shows the experimental SAXS data for a 200~mM concentration of ADOH in \ce{C6H12} taken from Ref. \cite{Facchin2017}.
Interestingly, the line is not characterised by a marked drop of the intensity as expected for regular spherical objects of well-defined radius.
Rather, at higher scattering vector $Q$ it decays slowly and without showing any particular feature, a behaviour compatible with the presence of irregular objects with no clear size.
This is in very good qualitative agreement with our simulations, which show the absence of a well-defined organization of the surfactant.
The molecular RDF, in particular, is characterised by a simple profile with a short-range peak and a fast decay (Figure~\ref{fig:gromacs_chex_struct}C).
Assuming a similar situation in the experiment, with an exponentially decaying RDF, the spectral SAXS line would then have the following form: 
\begin{equation}\label{eq:intensity_saxs}
    I (Q) = \frac{A}{1 +\xi^2 Q^2},
\end{equation}
where $\xi$ is the Ornstein-Zernike correlation length \cite{Barrat03}.
As can be observed in Figure~\ref{fig:saxs_fit}, such line shape fits well the experimental data. 
For comparison, we report the predicted SAXS spectrum obtained by Fourier transform of the RDF from the all-atom simulation, as well as its fitting by Eq.~\ref{eq:intensity_saxs}. 
Like for the diffusion coefficient, the agreement with the fit on the SAXS measurements is an indication that our computational model is well in line with the experimental findings.

\paragraph{Driving forces of ADOH interactions in \ce{C6H12}} 
The standard understanding of micelle formation is based on the hydrophobic effect, which is an entropy driven phenomenon.
By contrast, in \ce{C6H12} the formation of ADOH clusters is likely to be penalized entropically, due to the necessary confinement of TEG tails in the micellar core.
Rather, ADOH aggregates might be stabilized by the enthalpy gain associated to the electrostatic interactions occurring among the polar TEG segments when two TEG tails rest at close distance.
If so, the lower diffusion coefficient for ADOH would be explained in terms of molecular crowding, where the pathway of a freely diffusing ADOH is hindered by the interaction with other ADOH molecules.
This would have the effect of increasing the local viscosity of the medium, thus reducing diffusion even in the absence of a stable aggregate formation.

To verify this hypothesis, we ran additional all-atom simulations of ADOH in both water and \ce{C6H12} where all partial charges on the TEG segment of the amphiphile were set to zero, and replaced by an effective dipole in the final hydroxyl group mimicking the total dipole of the tail.
These artificial systems keep a coarse representation of the polarity of the TEG tail, while at the same time removing the quadrupolar charge distributions characteristic of the glycol ether moieties.
The presence of a finite dipole on the terminal OH also ensures a correct representation of the terminal H-bonding group, which was hypothesized to be crucial for ADOH aggregation in \ce{C6H12}.

\begin{figure} 
	\centering
    \includegraphics{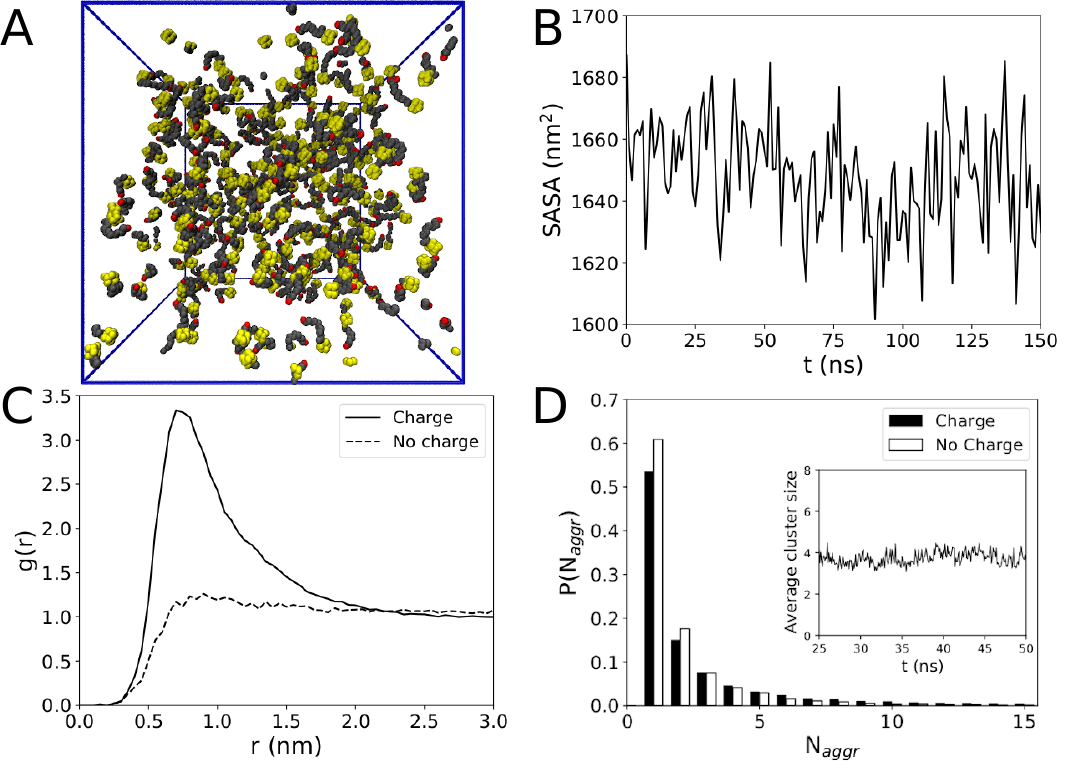}
	\caption{(A) All-atom simulation of 200 mM ADOH in C$_6$H$_{12}$ without partial charges. (B) SASA as a function of simulation time. (C) Comparison of RDF with and without partial charges. (D) Same as in (C) for the cluster size distribution.}
	\label{fig:chex_no_charge}
\end{figure}

Simulations in water, discussed in Section~\ref{subsec:water}, resulted in a faster formation of micelles compared to what was observed for the physical model (not shown).
In \ce{C6H12} the absence of quadrupolar charges on the TEG segment reduces even further the aggregation of ADOH, resulting in a net 30\% increase of its diffusion coefficient (see second column in Table \ref{tab:diffusion}).

The final snapshot of this system in \ce{C6H12} after \SI{150}{ns} is shown in Figure~\ref{fig:chex_no_charge}A, while the SASA is reported in Figure~\ref{fig:chex_no_charge}B.
Clear differences in the aggregation behaviour are evident by inspecting the RDF between ADOH molecules (Figure~\ref{fig:chex_no_charge}C).
In the absence of partial charges on the TEG segment, the RDF does not show a clear first-neighbour peak. 
On the contrary, the RDF reaches unity immediately at $r = \SI{1}{nm}$, which implies that each molecule is spatially uncorrelated from one another and that the distribution of surfactant in the sample is practically uniform.
The distribution of the aggregation number (Figure~\ref{fig:chex_no_charge}D) shows a marked difference with the distribution more skewed toward the monomer peak rather than larger aggregates, and an average cluster size of $N_{\text{aggr}}=4$ significantly smaller than the original model with all the partial charges (compare with the inset of Figure \ref{fig:gromacs_chex_struct}D).

\begin{figure}
	\centering
    \includegraphics{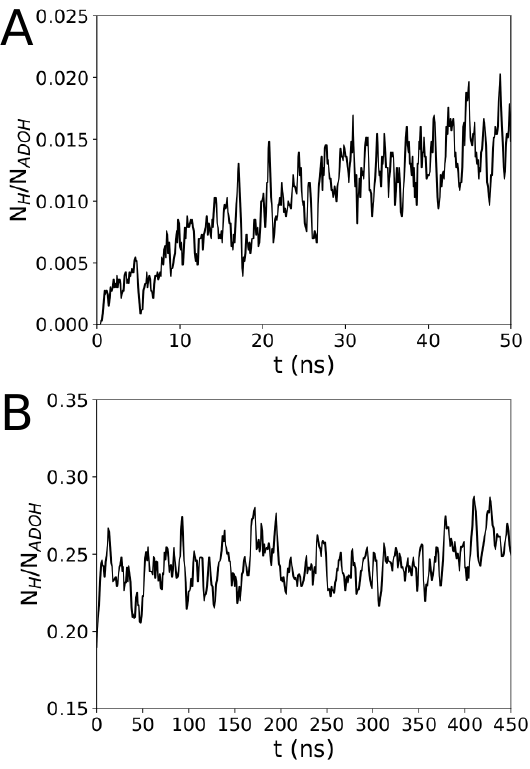}
	\caption{Number of hydrogen bonds between ADOH molecules as a function of time in all-atom simulations in (A) water and (B) \ce{C6H12}.}
	\label{fig:hbond}
\end{figure}

Formation of hydrogen bonding between the hydroxyl groups at the tails of ADOH was also putatively indicated as a possible driving force for ADOH aggregation in C$_6$H$_{12}$. 
Figure~\ref{fig:hbond} reports the number of hydrogen bonds per ADOH molecule ($N_H/N_{ADOH}$) as a function of time, in water and \ce{C6H12}.
As expected, the number of H-bonds between ADOH in water is marginal, as the strongly hydrophilic groups prefer to interact with the solvent (see Figure \ref{fig:hbond}A).
However, in \ce{C6H12} a significant fraction of ADOH is involved in H-bonds, with a very stable average $\approx 0.25$ H-bonds per molecule, as shown in Figure \ref{fig:hbond}B.
This number is anyway much smaller than the theoretical number of H-bonds expected in the sample if clustering between ADOH molecules was determined by H-bonding.
In fact, assuming that all clustered ADOH molecules would be involved in at least a single H-bond, and considering the average cluster size (see inset Figure~\ref{fig:gromacs_chex_struct}D), one would expect a value of $\sim$ 0.87 H-bonds per molecule, which is almost three times the one observed.
This finding confirms that while H-bonds between ADOH tails can contribute to the binding interactions, they themselves are not strong enough to constitute a significant driving force for ADOH aggregation.

Overall, simulations of the toy system without TEG charges confirm that cohesion of ADOH in \ce{C6H12} is an enthalpy-dominated phenomenon, mainly due to quadrupole-quadrupole interactions between the TEG polar regions. 
This is qualitatively different from the situation in water, where the collapse of the surfactant is determined by the entropy, and where the same qualitative effect could be obtained by lumping all partial charges of TEG into an effective dipole on the hydroxyl group at the end of the tail.
Our simulations in \ce{C6H12} indicate the presence of small, labile ADOH clusters rapidly forming and dissolving in the sample.
The presence of an appreciable density dishomogeneity in the absence of true stable aggregates is indicative of a mixture at supercritical conditions, confirmed by the shape of the SAXS profile, in qualitative agreement with those observed in other supercritical mixtures, like, for example, water/acetonitrile\cite{Nishikawa2002}.

\section{Conclusions}
\label{sec:conclusions}
We performed a detailed computational analysis of the self-assembly process of ADOH surfactants composed by a hydrophobic adamantane head and a TEG tail terminating with a \ce{OH} hydroxyl group. 
We have studied the mechanism driving the formation of aggregates in \ce{C6H12}, by comparing their sizes, distributions, interactions, as well as their diffusion coefficients with their experimental counterparts.\cite{Facchin2017}
In addition, we have also studied the same system in water providing a qualitative prediction for all the above quantities that could be experimentally tested.
By a detailed analysis of the results under different concentrations, we have also been able to estimate a value of \SI{13.5}{mM} for the CMC in water.

Our parallel analysis in water and \ce{C6H12} has allowed us to underpin the different forces driving the self-assembly in the two solvents.
In water, the conventional hydrophobic effect plays a major role in promoting the aggregation of the hydrophobic heads that are buried inside the micelles to avoid direct contact with water molecules. 
In \ce{C6H12}, an analogous {\it lipophobic} effect does not occur, and formation of well defined micelles with inverted structure was not observed. 
The lack of such an effect may be attributed to concomitant factors, including the fact that TEG chains are not as lipophobic as adamantane is hydrophobic, and that the conformational frustration of the flexible TEG moieties is insufficiently balanced by the solvent entropy release associated to the collapse of an aggregate.  

Our investigation highlighted the fact that the aggregation of ADOH into low-weight oligomers is sufficient to explain the decrease of diffusion coefficient experimentally observed, as well as the previously reported SAXS data. 
Formation of these oligomers is determined by short-range attraction of the electrostatic quadrupoles distributed over the TEG segment.
By contrast, in water these interaction are favoured by the low dielectric screening due to the weakly polarizable solvent.

However promising, ADOH in \ce{C6H12} appears to linger in a supercritical phase, without a clear collapse into well defined self-assembled aggregates.
Our study points to multiple possible routes that could produce significant steps toward such a goal.
Firstly, there is the need of decreasing the entropy of the free TEG chain to reduce the entropy loss upon confinement of the polar regions in the micellar core.
This could be achieved by narrowing the accessible conformational space of the TEG chain already in the monomer, for example, by adding a second TEG unit to the ADOH structure.
Additionally, it might be possible to chemically modify the TEG part, or to use other polymers having an even stronger polar character.
While these two options aim for raising the critical temperature of the systems, it is worth mentioning the even simpler idea of studying ADOH dissolved in low freezing point solvents.
This option should be considered in particular thinking to other thermodynamic regimes (high pressure, low temperature) that could be found in extraterrestrial environments.   
Synthetic effort in this direction may soon lead to the determination of surfactants having the ability of inverting their aggregation structure in solvents of radically different polarity.

\section{Supporting Information}
We provide a short movie for the all-atom simulation of ADOH in \ce{C6H12} in order to show the fast aggregation/disgregation process.

\section{Acknowledgement}
We are indebted to Paul Dupire and Emanuele Petretto for their help at the initial stage of the project, and to the authors of Ref.\cite{Facchin2017} for useful discussions. The authors would also like to thank Reidar Lund for his precious help with interpretation of SAXS data. 
This work was supported by the Research Council of Norway(RCN) through the CoE Hylleraas Centre for Quantum
Molecular Sciences (Grant number 262695), by the Norwegian Supercomputing Program (NOTUR) (Grant number NN4654K), by MIUR PRIN-COFIN2017 \textit{Soft Adaptive Networks} grant 2017Z55KCW , by Marie Curie Sklodowska-Curie Fellowship No. 894784 EMPHABIOSYS and by a Knight Chair to Prof. Jayanth Banavar at University of Oregon (T.S). The use of the SCSCF multiprocessor cluster at the Universit\`{a} Ca' Foscari Venezia is gratefully acknowledged. The authors would like to acknowledge networking support by the COST Action CA17139.


\end{document}